# Matching the photocurrent of perovskite/organic tandem solar modules by varying the cell width[†]

*José García Cerrillo[*], Andreas Distler, Fabio Matteocci, Karen Forberich, Michael Wagner, Robin Basu, Luigi Angelo Castriotta, Farshad Jafarzadeh, Francesca Brunetti, Fu Yang, Ning Li, Asiel Neftalí Corpus Mendoza, Aldo Di Carlo, Christoph J. Brabec[*], Hans-Joachim Egelhaaf*

J. García Cerrillo, Dr. A. Distler, R. Basu, Prof. N. Li, Dr. H.-J. Egelhaaf, Prof. C. J. Brabec
Institute of Materials for Electronics and Energy Technology (i-MEET)
Friedrich-Alexander-Universität Erlangen-Nürnberg (FAU)
Martensstraße 7, Erlangen, D-91058, Germany.
[*] E-Mail address: jose.garcia@fau.de.

Dr. F. Matteocci, Dr. L. A. Castriotta, F. Jafarzadeh, Dr. Francesca Brunetti, Prof. A. Di Carlo
Center for Hybrid and Organic Solar Energy (CHOSE)
University of Rome Tor Vergata
via del Politecnico 1
00133 Rome, Italy.

Dr. K. Forberich, M. Wagner, Prof. N. Li, Dr. H.-J. Egelhaaf, Prof. C. J. Brabec
Helmholtz Institute Erlangen-Nürnberg for Renewable Energy (HI ERN)
Forschungszentrum Jülich GmbH
Immerwahrstraße 2, Erlangen, D-91058, Germany.
[*] E-mail address: christoph.brabec@fau.de.

Dr. F. Yang
Laboratory of Advanced Optoelectronic Materials
College of Chemistry
Chemical Engineering and Materials Science
Soochow University
Suzhou 215123, China.

Prof. N. Li
State Key Laboratory of Luminescent Materials and Devices
Institute of Polymer Optoelectronic Materials and Devices
School of Materials Science and Engineering
South China University of Technology
Guangzhou 510640, China.

Dr. Asiel Neftalí Corpus Mendoza
Instituto de Energías Renovables
Universidad Nacional Autónoma de México
Priv. Xochicalco S/N, Temixco, Morelos, 62580 México.

Prof. A. Di Carlo
Institute of Structure of Matter (ISM)
National Research Council (CNR)
via del Fosso di Cavaliere 100, 00133 Rome, Italy.

J. G. C., A. D. and F. M. contributed equally.

**Abstract**

Photocurrent matching in conventional monolithic tandem solar cells is achieved by choosing semiconductors with complementary absorption spectra and by carefully adjusting the optical properties of the complete top and bottom stacks. However, for thin film photovoltaic technologies at the module level, another design variable significantly alleviates the task of photocurrent matching, namely the cell width, whose modification can be readily realized by the adjustment of the module layout. Herein we demonstrate this concept at the experimental level for the first time for a 2T-mechanically stacked perovskite (FAPbBr$_3$)/organic (PM6:Y6:PCBM) tandem mini-module, an unprecedented approach for these emergent photovoltaic technologies fabricated in an independent manner. An excellent $I_{sc}$ matching is achieved by tuning the cell widths of the perovskite and organic modules to 7.22 mm ($PCE_{PVKT\text{-}mod}$= 6.69%) and 3.19 mm ($PCE_{OPV\text{-}mod}$= 12.46%), respectively, leading to a champion efficiency of 14.94% for the tandem module interconnected in series with an aperture area of 20.25 cm$^2$. Rather than demonstrating high efficiencies at the level of small lab cells, our successful experimental proof-of-concept at the module level proves to be particularly useful to couple devices with non-complementary semiconductors, either in series or in parallel electrical connection, hence overcoming the limitations imposed by the monolithic structure.

## 1. Introduction

More than 13 years after the first report of the application of perovskites in mesoscopic solar cells,[1] the photovoltaic technology based on this groundbreaking material has approached levels of maturity that may allow their introduction into the market of renewable energies in the short- to mid-term.[2,3] The ever-growing interest in this particular kind of hybrid organic-inorganic semiconductor material has been firmly grounded on several physical and chemical properties, such as strong light absorption,[4] long charge carrier diffusion lengths,[5] bandgap tunability,[6] distinctively high defect tolerance,[7] and others. Furthermore, the possibility to deposit perovskite thin films over large areas by a wide variety of relatively low-cost solution-based methods, like slot-die coating,[8,9] spray coating[10,11] and doctor blading,[12,13,14] makes them considerably attractive from an industrial perspective.[15]

Currently, state-of-the-art perovskite and silicon solar cells have achieved outstanding power conversion efficiencies of 26.0% [16,17] and 26.81%, [17,18] respectively. However, in order to have the possibility of fabricating highly efficient solar photovoltaic devices with wide absorption spectra and minimized charge carrier thermalization losses, perovskites have been integrated with silicon solar cells in *tandem* devices to obtain unprecedented efficiencies as high as 33.7%.[16] Their coupling is usually done in a 2-terminal interconnection (2T), where the photocurrents generated by both sub-cells must be equal so that their voltages add up to deliver the highest possible power. In principle, a 4-terminal (4T) tandem configuration would also be possible, in which the total efficiency equals the sum of the individual efficiencies delivered by the top and bottom cells without restrictions on either $I_{sc}$ or $V_{oc}$ matching.

However, it entails the need to employ additional wirings and inverters that would lead to increased material expenditures [19,20].

Although perovskite/silicon tandem solar cells are the frontrunners for commercialization in the short- to mid-terms, it is also possible to combine perovskites with other semiconductors like copper-indium-gallium selenide (CIGS)[21,22,23,24], gallium arsenide[25], perovskites[26,27,28,29], and organic materials[30,31,32,33]. The combination of a wide bandgap perovskite with a narrow bandgap organic module opens new fields of application, as it will allow the design of tandem devices with high average visual transmission, due to the excitonic nature of organic semiconductors[34].

In the conventional, series-connected 2T monolithic architecture the top and bottom sub-cells are fabricated using the same lateral layout, being joined by an interconnecting layer where the carriers of opposite charge recombine to maintain the electrical neutrality within the device[3]. In this case, at least two independent problems arise: 1) depositing the top module may damage one or several layers of the bottom module unless an impermeable indium-tin oxide (ITO) layer and/or a buffer film is employed, such as $SnO_2$ deposited by atomic layer deposition (ALD), due to the polar solvents used to process the perovskite film of the upper module[32,35], and 2) as the cells in both modules have exactly the same areas, the only parameter available for matching the currents of the two sub-modules are the thicknesses of the respective active layers. This limits the resulting tandem efficiency, especially for large differences in the bandgaps of the top and bottom active layers, as is the case for tandem modules designed for providing a high average visible transmittance (AVT). Furthermore, in real world applications[36,37], the daily variations of light intensity, incidence angle and temperatures in different regions make it even more difficult to keep the $J_{sc}$ matching required to obtain the maximum performance.

We demonstrate herein how both problems can be eliminated by fabricating perovskite and organic sub-modules individually and stacking them subsequently. Based on earlier reports theorizing the possibility to adjust the cell size or number of cells to match the currents of perovskite and silicon or CIGS modules [19,38] as well as theoretical models to forecast the performance of perovskite/silicon tandems under different outdoor conditions, [39,40] we make use of the linear relationship between the short circuit current, $I_{sc}$, and the active areas of the individual cells of the sub-modules to match the photocurrents delivered by a semitransparent wide-bandgap perovskite module based on FAPbBr$_3$ ($E_g$= 2.28 eV) to that of an opaque organic module based on the ternary system PM6:Y6:PCBM ($E_g$= 1.39 eV). If the cell areas in both sub-modules were identical, as would be the case for monolithic tandem modules, the large difference in the bandgaps of these semiconductors would impede the realization of efficient tandems in a monolithic architecture due to the significant photocurrent mismatch, the tandem $I_{sc}$ being limited by the perovskite sub-module.

However, by fabricating the sub-modules independently and connecting them in series, we accomplish an excellent $I_{sc}$ matching by combining a 6-cell top perovskite and a 13-cell bottom organic mini-module, both having an aperture area of 20.25 cm$^2$ and cell widths of 7.22 mm and 3.19 mm, respectively, thus leading to a champion tandem efficiency of 14.94%. Furthermore, the similarity of the $V_{oc}$ of these mini-modules enables their connection in parallel, leading to an efficiency of 14.11%, which is also higher than the individual sub-module efficiencies. To the best of our knowledge, these results represent the first experimental demonstration of successful $I_{sc}$ matching of thin film photovoltaic modules fabricated in separate production lines, a unique advantage of the mechanically-stacked tandems that lead to higher efficiencies than the monolithic structure for non-complementary semiconductors.

The concept demonstrated herein can be extended to match the $I_{sc}$ or the $V_{oc}$ of bigger modules with already optimized deposition conditions, thus circumventing the need to carefully optimize each of the layers comprising a monolithic tandem. Moreover, the replacement of interconnecting layers with simple wire-based connections offers the possibility to simplify the manufacturing process, reduce the fabrication costs, and decrease the optical and electric losses associated with interconnecting layers. Hence, this strategy opens new horizons to simplify the fabrication of tandem modules, and at the same time enables a wider selection of light absorbing materials.

## 2. Results and Discussion

The concept of individual cell widths in bottom and top modules is explained as follows. The general structure of our 2T-interconnected perovskite/organic tandem solar modules is depicted in **Figure 1a**, which contrasts with the conventional monolithic architecture shown in **Figure 1b**. The photocurrent generated in the tandem module is collected from the terminals of both sub-modules, which are interconnected by means of electrical wiring. The $V_{oc}$ of the interconnected tandem equals the sum of the individual $V_{oc}$ of the perovskite ($V_{oc,PVK}$) and organic ($V_{oc,f-OPV}$) modules and, in order to reach the highest possible efficiency, the short-circuit current generated by the top semitransparent perovskite module ($I_{sc,PVK}$) must equal the $I_{sc}$ of the organic module ($I_{sc,f-OPV}$) exposed to sunlight filtered by the top module.

Here, the *filtering* effect imparted to the bottom organic module refers to two aspects: 1) the absorption of the short-wavelength range of the electromagnetic spectrum by the top perovskite module, and 2) the decrease in the light intensity resulting from the absorption and reflection occurring in the top perovskite module.

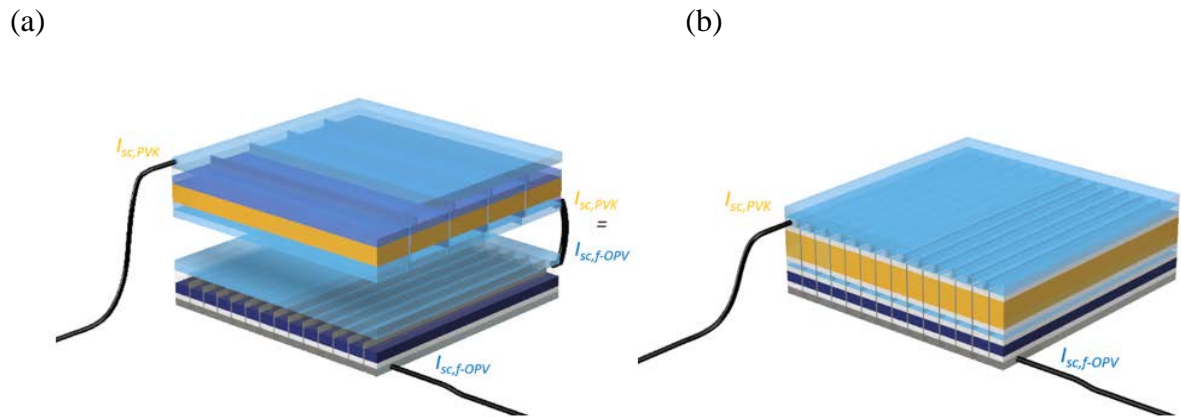

**Figure 1**. Perovskite/organic tandem solar modules. (a) 2T mechanically-stacked architecture with semitransparent electrodes in the upper perovskite module, and (b) monolithic architecture with a recombination layer.

According to Kirchhoff's laws, the electrical current flowing through each of the elements connected in series is the same. In the case of an ideal solar module, i.e. with infinitely high shunt resistance and low series resistance, the total current is the same as the current photogenerated in each of its constituting cells. A linear relationship exists between the photocurrent generated in a solar cell and its active area, which in turn varies linearly with the cell width for a constant cell length (*A*= width x length). From this it follows that the total current delivered by a module can be adjusted by varying the width of its constituting cells, hence making it possible to simplify the process of current matching of two modules to fabricate a 2T mechanically stacked tandem module. This additional degree of freedom is unique to the mechanically stacked tandem architecture, an advantage that can be conveniently applied at the industrial level.

The electric current $I_{sc,PVK}$ of the perovskite sub-module is obtained as the product of the short circuit current density ($J_{sc,PVK}$), which increases with the thickness of the $FAPbBr_3$ film, and the active area ($A_{ac,PVK}$) of its individual cells. Since all the cells have the same length, the active area is directly proportional to the cell width ($w_{ac,PVK}$). Likewise, the electric current delivered by the bottom organic module depends on the light transmitted by the perovskite module, the thickness of the PM6:Y6:PCBM film ($d_{OPV}$) and the cell width ($w_{ac,f-OPV}$). The intensity of light transmitted to the bottom organic module is also reduced by the parasitic absorption and reflection across the layers comprising the top perovskite module, especially by the FTO and ITO films employed as the top and bottom electrodes, respectively, as observed in Figure 1a.

For a given semiconductor employed in the bottom cell of a monolithic tandem, the condition necessary to attain the highest possible efficiency is to choose a wide bandgap cell that absorbs high-energy photons more efficiently than the bottom semiconductor, hence reducing the excess energy lost as heat (e.g. thermalization losses) in the latter for $E > E_g$. In a working device, this is reflected by a maximization of both the tandem $J_{sc}$ and $V_{oc}$, which is only achieved by a specific combination of wide and narrow bandgaps. The tandem is thus designed to have semiconductors with *complementary* bandgaps. For example, the specific wide bandgap that complements the bandgap of silicon (1.12 eV) is within the range 1.68-1.70 eV, as demonstrated by the highly efficient tandems fabricated by Mariotti et al.[41].

As shown in **Figure S1**, the external quantum efficiency (EQE) spectra of semitransparent perovskite and opaque organic cells based on FAPbBr$_3$ and PM6:Y6:PCBM display absorption onsets at 544 nm ($E_g$= 2.28 eV) and 892 nm ($E_g$= 1.39 eV), respectively. According to efficiency maps previously reported for a 2T interconnection[19], the complementary perovskite/OPV bandgap pair for FAPbBr$_3$ would be 2.28 eV/1.77 eV and 1.93 eV/1.39 eV for PM6:Y6:PCBM, which means that our tandem fabricated with these two active layers will have *non-complementary* bandgaps. Nonetheless, given a certain film thickness of the FAPbBr$_3$ layer in the semitransparent perovskite module, it is possible to match its $I_{sc}$ to that of the organic PM6:Y6:PCBM module by adjusting the cell widths, as is demonstrated in the following with optical simulations and experiments.

## 2.1. Optical simulations

In order to assess the design flexibility enabled by the adjustment of the cell width to match the module photocurrents, we perform optical simulations based on the transfer matrix method. The optical constants of the films comprising the individual modules are given in **Figure S2**, and the full details of the method are provided in the Supporting Information.

Two cases are considered to attain the current matching, 1) monolithic tandem cell, where only the variation of the film thickness or optical density (*OD*) is possible (same active area for both cells, as seen in **Figure 2a** and **2b**), and 2) our tandem concept, where the cell area (i.e. width) of the bottom organic cell is adjusted alongside the *OD* of the perovskite and organic films to achieve the desired $I_{sc}$ matching. **Figure S3a** illustrates the *OD* spectra of the FAPbBr$_3$ and PM6:Y6:PCBM (at 620 nm) films with thicknesses of 150 nm and 110 nm, respectively and **Figure S3b** illustrates the $J_{sc}$ of the perovskite and organic cells as a function of the optical density of their respective active layers.

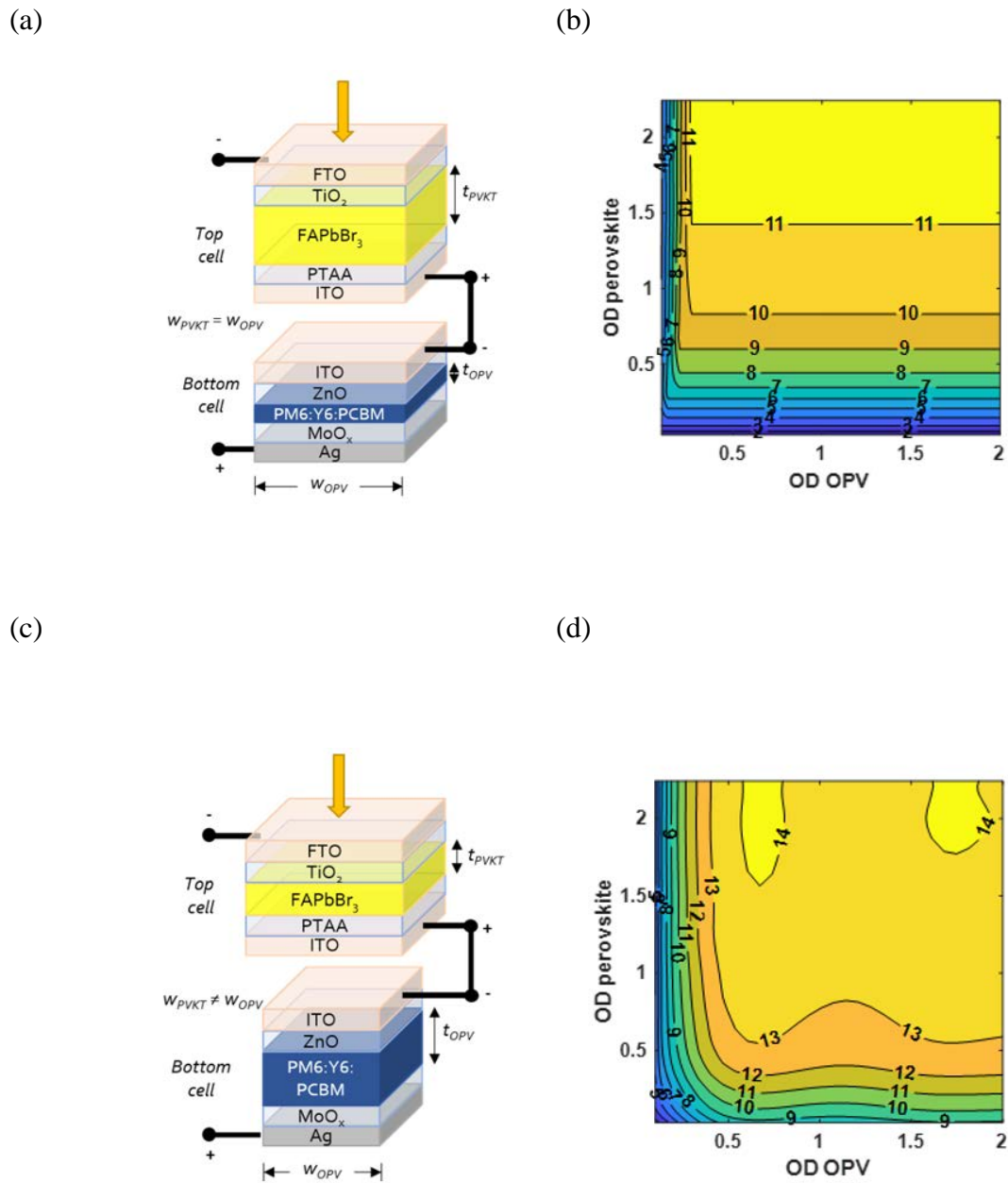

**Figure 2.** (a) Tandem solar device with top perovskite and bottom organic cells having equal active areas ($w_{PVKT} = w_{OPV}$), (b) 2T-tandem efficiency for equal cell dimensions as a function of the film optical density, (c) Tandem device with varying organic cell width to achieve current matching for fixed perovskite cell dimensions ($w_{PVKT} \neq w_{OPV}$), and (d) the corresponding 2T-tandem efficiency map. The efficiencies were calculated by considering the lowest $I_{sc}$, the sum of the individual $V_{oc}$ and the fill factor was calculated by considering the resistance losses as a function of the cell width. Full details are shown in the Supporting Information.

In the case of a monolithic interconnection, the cell areas of both cells are equal (as shown in Figure 2a), therefore the ratio $I_{sc,f\text{-}OPV}/I_{sc,PVK}$ equals the ratio $J_{sc,f\text{-}OPV}/J_{sc,PVK}$. This is the case encountered in all of the tandem solar cells and modules reported so far. Consequently, as observed in Figure 2b, a perfect $I_{sc}$ matching and a maximum tandem efficiency marginally above 11% can be obtained for an organic film with $OD_{620\,nm}$ larger than 0.27, which corresponds to a film thicknesses higher than 53 nm that should be combined with a FAPbBr$_3$ film of $OD$ higher than 1.42. The deposition of such ~50 nm-thick PM6:Y6:PCBM films would severely compromise the performance of the organic module due to the increased risk of formation of pinholes, especially when larger substrate areas are considered.

At $OD_{620\,nm}$ lower than 0.27 ($t_{OPV} < 53$ nm), the cells of the organic module would deliver a current lower than that of the perovskite module, that is, $J_{sc,f\text{-}OPV}/J_{sc,PVK} < 1$ (see **Figure S4a**) hence deviating from the ideal condition necessary for a series interconnection. However, if the $OD_{620\,nm}$ of the organic film is significantly larger, the photocurrent delivered by the organic module will be larger than that of the top module regardless of the perovskite film thickness ($J_{sc,f\text{-}OPV}/J_{sc,PVK} > 1$). In this case, the performance of the tandem module is always limited by the perovskite module regardless of the thickness of the PM6:Y6:PCBM active layer and, therefore, the highest attainable efficiency remains at the 11% threshold. These findings highlight one of the major shortcomings of monolithic tandems, which are built on the same substrate and therefore confine the top and bottom cells to the same active area. The maximum performance that can be attained with this specific perovskite/organic pair is considerably limited by the restrictive set of processing variables, i.e. only the film thicknesses of the layers within the stack can be adjusted.

In stark contrast, if the cell width of the organic sub-module is adjusted alongside the $OD$ of both active layers, as shown in **Figures 2c** and **2d**, the processing window of the bottom organic layer is widened to reach higher efficiencies. Importantly, for different cell areas or widths of the perovskite and organic modules, the ratio $I_{sc,f-OPV}/I_{sc,PVK}$ does not equal the ratio $J_{sc,f-OPV}/J_{sc,PVK}$, which is the distinctive case proposed in this work. Nevertheless, if the perovskite and organic modules have the same module area and deliver the same electric current, then the requirement for a series interconnection remains: $[J_{sc,PVK}]_{module} = [J_{sc,f-OPV}]_{module}$, while at the same time $[J_{sc,PVK}]_{cell} \neq [J_{sc,f-OPV}]_{cell}$.

Our optical simulations in Figure 2d demonstrate that tandem efficiencies higher than 14% are obtained for modules having organic films with optical densities ranging between the intervals 0.58-0.80 and 1.62-1.87, which correspond to film thicknesses between 126-179 nm and 373-433 nm, respectively. Such an organic film would have to be combined with a perovskite film with *OD* higher than 1.56. Preferably, the organic film thickness should lie within the first range to avoid recombination losses and their resultant lower fill factor. In fact, the range found in our simulations is close to the PM6:Y6:PCBM film thickness (~110 nm) leading to organic solar modules with record efficiencies, as reported by Distler et al.[42]. In the present case of adjustable cell widths, the efficiency of the tandem is no longer limited by the $I_{sc}$ of the perovskite module, since the $I_{sc}$ of the organic module can be conveniently tuned to attain the ideal current matching. **Figure S4b** displays the cell width of the OPV module necessary to achieve the current matching with the semitransparent perovskite module, while the **Figure S5** displays the losses related to the geometric fill factor, resistance and efficiency resulting from an increasing organic cell width.

The additional degree of freedom demonstrated herein at the module level, i.e. photocurrent matching via the cell adjustment, will alleviate the design of mechanically stacked tandems that do not impose the necessity to deposit an energy- and time-consuming recombination layer. Furthermore, this strategy enables the matching of modules with non-complementary semiconductors, being solely based on a geometrical optimization related to module patterning without varying the delicate equilibrium of the cell stack composition, sequence and thickness as performed in a conventional 2T tandem design. Based on our optical simulations, in the following we provide experimental evidence that validates the strategy proposed in this work.

**2.2. Perovskite/organic tandem solar modules**

*2.2.1. Individual module design*

In order to prove the concept described above, different tandem modules were fabricated, consisting of a 6-cell semitransparent wide-bandgap FAPbBr$_3$ perovskite module and PM6:Y6:PCBM organic solar modules with different cell widths corresponding to 12 and 13 cells connected in series. The FAPbBr$_3$ modules have high average visible transmittance (*AVT* ∼60%) and high geometric fill factor (>97%). Likewise, the PM6:Y6:PCBM modules show outstanding *FF* values above 73% and *GFF* above 96%[42]. **Table 1** summarizes the geometrical characteristics of the perovskite and organic modules and **Figure 3** displays photographs of a perovskite module and the tandem stack during the measurements. Further experimental details are provided in the Supporting Information. The bandgaps of the perovskite and organic active layers were calculated from their respective EQE spectra as reported by Almora et al. [43]

**Table 1.** Geometrical parameters of the semitransparent (ST) perovskite and organic solar modules used for the fabrication of tandems.

| Module | $E_g$ (eV) | $n_{cells}$ | $A_{ac,cell}$ (cm$^2$) | $w_{ac,cell}$ (mm) | $w_{IC}$ (µm) | $A_{T, active}$ (cm$^2$) | $A_{T, inactive}$ (cm$^2$) | $A_{total}$ (cm$^2$) | GFF (%) |
|---|---|---|---|---|---|---|---|---|---|
| ST-Perovskite | 2.28 | 6 | 3.32 | 7.38 | 160 | 19.93 | 0.44 | 20.37 | 97.83 |
| PM6:Y6:PCBM | 1.39 | 12 | 1.63 | 3.63 | 150 | 19.61 | 0.64 | 20.25 | 96.82 |
|  |  | 13 | 1.43 | 3.19 | 150 | 18.65 | 1.60 | 20.25 | 92.08 |

$A_{ac,cell}$ and $w_{ac,cell}$- active area and width of each cell, $w_{IC}$- width of the interconnects, $A_{T,active}$- total active area, $A_{T,inactive}$- total inactive area, $A_{total}$- total module (or aperture) area, *GFF*- Geometric Fill Factor.

(a) 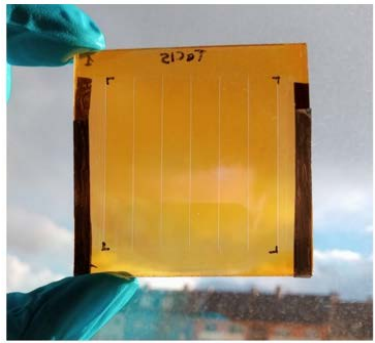   (b) 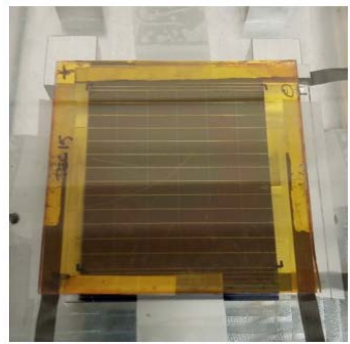

(c) 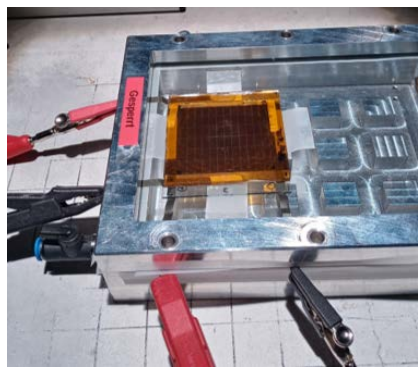

**Figure 3.** (a) Photograph of a semitransparent, FAPbBr$_3$-based perovskite solar module, (b) perovskite/organic tandem solar module with the interconnect lines arranged in a perpendicular orientation, and (c) measurement set up under 100 mW/cm$^2$ illumination and continuous N$_2$ flow.

*2.2.2. Cell width of the organic solar modules*

**Table 2** and **Table S1** show the highest and average photovoltaic parameters of the tandems fabricated with organic solar modules having 12 and 13 cells. The experimental results of the tandems connected in series are compared to the parameters expected for infinitely high shunt resistances, namely the lowest $I_{sc}$ between the perovskite and filtered organic module, the sum of the individual $V_{oc}$ and the average $FF$. The power conversion efficiency shown in Table 2 refers to the total aperture area, i.e. 20.25 cm$^2$, instead of the total active area. **Figure 4** displays the current-voltage characteristics of the individual modules as well as those of the tandems connected in series and parallel.

The first aspects to notice are the difference in the $I_{sc}$ of the semitransparent perovskite and filtered organic modules as well as the effect of the relative current mismatch on the tandem efficiency. As depicted in **Figure 4a** and in Table 2, the $I_{sc}$ of the organic module with 3.63 mm-wide (twelve) cells is larger than the $I_{sc}$ of the perovskite module by a percentual mismatch (% $\Delta I_{sc}$) of 8.92% with respect to the perovskite device. That is, the $I_{sc}$ of the top perovskite module limits the $I_{sc}$ of the tandem connected in series, although the latter (21.51 mA) is higher than the expected value of 20.40 mA.

On the other hand, in accordance with **Figure 4b**, the $I_{sc}$ of the organic module containing 3.19 mm-wide (thirteen) solar cells is only $\Delta I_{sc}$= 2.61% lower than the $I_{sc}$ of the perovskite module. That is, there is a 3.42-fold improvement of the $I_{sc}$ matching by reducing the cell width from 3.63 mm to 3.19 mm, which has the advantage of having one more cell to improve the tandem $V_{oc}$ and also a higher $FF$ due to a smaller current flowing through the ITO electrode. The fine tuning of the sub-module $I_{sc}$ attainable with this strategy is unique to the mechanically stacked

2T architecture, since for a monolithic tandem only the layer thicknesses of the active layers can be adjusted to get the photocurrent matching.

**Table 2**. Summary of results of the perovskite/organic tandem solar modules with varying cell width or number of cells in the organic module. The numbers on blue and light orange backgrounds are relevant to the series and parallel interconnection, respectively.

| Module | | | $J_{sc}$ (mA/cm$^2$) | $I_{sc}$ (mA) | $V_{oc}$ (V) | $I_{mpp}$ (mA) | $V_{mpp}$ (V) | $P_{mpp}$ (mW) | FF (%) | $PCE_{mod}$ (%) |
|---|---|---|---|---|---|---|---|---|---|---|
| **OPV with 12 cells (3.63 mm)** | | | | | | | | | | |
| ST-Perovskite | | | 6.15 | 20.40 | 9.17 | 17.39 | 7.16 | 124.51 | 66.56 | **6.11** |
| Organic (filtered) | | | 13.60 | 22.22 | 8.87 | 19.28 | 6.78 | 130.67 | 66.30 | 6.45 |
| Organic | | | 22.42 | 36.64 | 9.57 | 31.48 | 7.06 | 222.38 | 63.44 | **10.98** |
| Tandem | Series | Experimental | 1.06 | 21.51 | 18.12 | 18.29 | 14.12 | 258.22 | 66.26 | **12.75** |
| | | Expected | 1.01 | 20.40 | 18.04 | | | 244.47 | 66.43 | 12.07 |
| | Parallel | Experimental | 2.12 | 43.02 | 9.01 | 37.20 | 6.87 | 255.67 | 65.96 | **12.63** |
| | | Expected | 2.10 | 42.59 | 9.01 | | | 254.92 | 66.43 | 12.59 |
| **OPV with 13 cells (3.19 mm)** | | | | | | | | | | |
| ST-Perovskite | | | 6.46 | 21.46 | 9.20 | 18.08 | 7.16 | 129.45 | 65.55 | **6.36** |
| Organic (filtered) | | | 14.57 | 20.90 | 10.62 | 17.80 | 8.57 | 152.54 | 68.68 | 7.53 |
| Organic | | | 22.92 | 32.87 | 10.79 | 28.43 | 8.88 | 252.50 | 71.20 | **12.47** |
| Tandem | Series | Experimental | 1.04 | 21.07 | 20.05 | 17.93 | 16.20 | 290.48 | 68.76 | **14.34** |
| | | Expected | 1.03 | 20.90 | 19.83 | | | 278.16 | 67.12 | 13.74 |
| | Parallel | Experimental | 2.11 | 42.79 | 10.10 | 36.24 | 7.80 | 282.69 | 65.41 | **13.96** |
| | | Expected | 2.09 | 42.37 | 9.93 | | | 282.40 | 67.12 | 13.95 |

Noteworthy, we observed that the tandem $I_{sc}$ and *FF* are very close to the average of both modules. This effect is due to a relatively low shunt resistance of the module with the highest $I_{sc}$, which leads to current leakage that increases the current that can be delivered by the tandem module. To demonstrate this observation, we simulated the behavior of a perovskite/organic tandem module connected in series in LTspice, whose equivalent electric circuit is displayed in **Figure S6a** with the corresponding parameters summarized in **Table S2a** and **S2b**. For the individual sub-modules, a one-diode model was employed including elements corresponding to series ($R_s$) and shunt resistances ($R_{sh}$)[44]. When $R_{sh}$ is relatively low for both sub-modules, the tandem $I_{sc}$ tends to locate in between the individual $I_{sc}$, which fully supports our experimental results. However, for infinitely high (or high enough) $R_{sh}$, the tandem $I_{sc}$ equals the lowest $I_{sc}$, as expected from Kirchhoff's law (see **Figure S6b**). The latter effect on the short circuit current has been observed in previous reports, as shown in **Table S3**.

Additionally, in our perovskite/organic tandem configuration we noticed a negligible contribution in the $I_{sc}$ of the perovskite module from the light reflected at the surface of the substrate of the organic module and its silver electrodes. As observed in **Table S4**, the $J_{sc,cell}$ of a semitransparent FAPbBr$_3$ module had a minor increase from 6.45 mA/cm$^2$ to 6.66 mA/cm$^2$, which precludes any significant contribution of the back reflection from the organic module.

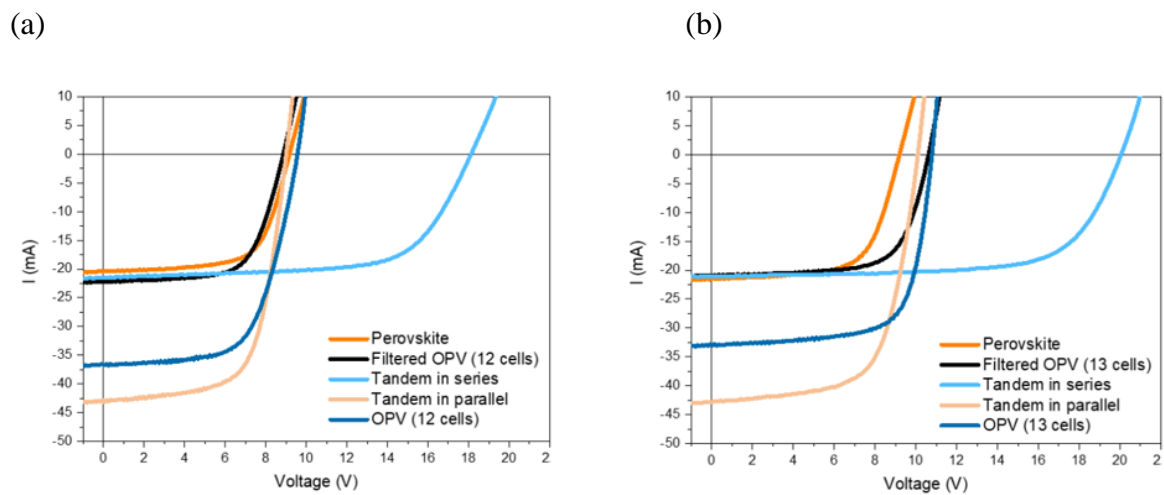

**Figure 4.** Current-voltage characteristics of tandem perovskite/organic solar modules connected in series and parallel. The organic solar modules have cell widths of (a) 3.63 mm (12 cells) and (b) 3.19 mm (13 cells).

For the two types of organic modules, the $V_{oc}$ of the tandem in series is slightly higher than the expected value, pointing to small (e.g. temperature-induced) variations in the individual $V_{oc}$ values during the measurements. Nevertheless, the high tandem $V_{oc}$ attained during our measurements proves the excellent interconnection in series without the need to employ a recombination layer, which is required in 2T monolithic tandems to keep the electric neutrality within the device. As seen in Figure 3c, our mechanically stacked modules were simply connected with electrical clamps, a straightforward strategy that would facilitate the fabrication of tandems at a large scale.

These results support the known fact that the photocurrent from devices forming a tandem in series connection must have a mismatch as small as possible, which then as a consequence maximizes the tandem $V_{oc}$ and the *FF*, as seen in the Table 2. We demonstrate herein that this is readily doable by making the appropriate modifications in the module layout.

The efficiency of the tandems employing 12 cell- and 13 cell-organic solar modules are equal to 12.75% and 14.34%, respectively, which is a noticeable improvement with respect to the standalone perovskite and organic modules that is in line with our optical simulations. The device performance of a tandem fabricated with the 13-cell module was further raised to 14.82% when a FAPbBr$_3$ module with slightly higher $V_{oc}$ and *FF* was employed, as seen in **Figure S7**, **Table S5** and **Table S6**. Moreover, a maximum tandem efficiency of 14.94% was recorded when the voltage step size of the I-V scan increased from 50 mV to 176 mV, hence making it possible to deliver a power slightly higher than 300 mW, as shown in the **Figure S8**. In an attempt to increase the *FF* of the perovskite/organic tandems, we connected the 13-cell organic module to a perovskite module deposited on an FTO substrate with lower sheet resistance (Tec 7), which effectively led to a tandem *FF* of 71.16% as shown in **Figure S9**,

**Table S7** and **Table S8**. Nevertheless, this increase was counterbalanced by a lower tandem $I_{sc}$ that resulted from a higher parasitic light absorption from the thicker FTO film.

Regarding the connection in parallel, we observed that it is also possible to match the $V_{oc}$ of the individual modules by adjusting their cell width. The $V_{oc}$ of the tandem (9.01 V) with a 12-cell organic module was almost exactly the average of the perovskite and filtered organic modules, while its $I_{sc}$ and *FF* were equal to 43.02 mA (0.96% higher than the sum of the individual perovskite and filtered organic $I_{sc}$) and 65.96%, respectively, leading to an efficiency of 12.63%. Although the tandem with the 13-cell module has a larger $V_{oc}$ mismatch, its higher efficiency of 13.96% is largely the result of an additional cell that overcompensates the slightly lower $I_{sc}$ and *FF* compared to those of the tandem with the 12-cell organic module. In the case of the 13-cell module, the tandem $V_{oc}$ of 10.10 V is closer to the highest $V_{oc}$, which is delivered by the filtered organic module. This is due to the series resistance of the individual modules and can be rationalized by analyzing their *I-V* curves in the vicinity of their respective $V_{oc}$, as shown in **Figure S10**. Our results regarding the parallel interconnection provide experimental evidence that would help in the design of efficient 2T-mechanically stacked tandem modules with reduced sensitivity to sudden changes in the radiation intensity or temperature, which in turn drive away the $I_{sc}$ matching. This is one of the known shortcomings of the series connection in the unpredictable conditions inherent in real-world applications[37]. Although we observe that the tandems connected in series tend to have a higher power conversion efficiency than the parallel configuration in the static, controlled illumination and temperature in the laboratory, it would be necessary to conduct a similar study outdoors to evaluate with certainty which configuration is the best for a determined application. For example, a parallel configuration could be preferred in partial shading conditions or in climates where non-uniform illumination is prevalent, which would allow the current generated by the illuminated cells to bypass the shaded cells. Further studies may also consider the implementation of encapsulating materials to compare the performance of protected and unprotected tandem modules against multiple environmental factors such as moisture, oxygen, heat, and light.

## 3. Conclusion

Based on simulations and experiments, we have proven that it is possible to fabricate perovskite/organic tandem modules with matched photocurrents and voltages via the adjustment of their cell width. This concept allows the independent fabrication of each sub-module, which avoids the risk of damaging the bottom module during the deposition of the top module, eliminates the requirement to deposit a recombination layer and adds a degree of freedom to the adjustment of the film thickness or optical density. To experimentally demonstrate this strategy, we combined a semitransparent, wide-bandgap $FAPbBr_3$ module with a narrow-bandgap PM6:Y6:PCBM module having efficiencies of 6.69% and 12.46%, respectively, to obtain a tandem module with efficiency of 14.94% on an aperture area of 20.25 $cm^2$. The focus of this contribution has been placed on the successful proof-of-concept of matching the $I_{sc}$ and $V_{oc}$ of thin film photovoltaic modules rather than the attainment of high efficiencies, which is commonplace for tandem solar cells with small active areas ($\leq 1.0$ $cm^2$) and complementary bandgaps. In contrast to the geometrically fixed monolithic structure, our findings show that adjusting the cell width for the development of mechanically-stacked tandem modules lead to higher efficiencies if non-complementary semiconductors are at hand. The enhanced design flexibility and independent fabrication enabled by this concept opens new opportunities in the design of semitransparent tandem modules for application in windows or greenhouses, whose visual transmittance can be conveniently adjusted by using a wider range of semiconductors.

**Experimental Section**

*Semitransparent wide-bandgap perovskite modules*

*Materials*

Anhydrous dimethylsulfoxide (DMSO) (99.9%), Ethyl acetate (99%, anhydrous), Titanium diisopropoxide (bis-acetylacetonate), Lithium bis(trifluoromethanesulfonyl)imide salt (Li-TFSI) (99.95%), 4-tert-butyl pyridine (tBP, 96%), Niobium ethoxide was purchased from Sigma Aldrich. $PbBr_2$ (99.99%) was purchased by TCI Chemical. Formamidinium bromide (FABr, 99.98%), neo-Pentylammonium chloride, iso-Pentylammonium chloride were purchased from Great Cell Materials. PTAA (10kDa) was purchased from Solaris Chem. Tin(IV) oxide, 15% in $H_2O$ colloidal dispersion was purchased from Alfa Aesar.

*Device Fabrication*

Transparent perovskite solar modules have been fabricated using patterned FTO substrates (2.2 mm-thick TEC15 Pilkington, 15 Ω/square) on 60x60 $mm^2$ substrate area by using picosecond pulsed laser (Wophotonics, Yb:KGW, λ = 355 nm, 5 ps, pulsed at 2000 kHz with a fluence per pulse for P1 lines equal to 40.85 mJ $cm^{-2}$). The patterned substrates were cleaned in an ultrasonic bath, using de-ionized water and soap and finally IPA (10 min for each cleaning step). The 40 nm-thick $TiO_2$ compact layer (c-$TiO_2$) is deposited by spray pyrolysis at 460°C starting from a precursor solution made up of titanium diisopropoxide bis(acetylacetonate), acetylacetone and ethanol in a 3:2:45 volume ratio. 2% Nb doping is made by adding Niobium Ethoxide in the precursor solution. $SnO_2$ nanoparticle-based ink at 1:20 v/v in deionized water is deposited on Nb:TiO$_2$ coated substrate using spin coating at 3000 rpm for 20s. Then, the substrates are annealed at 120°C for 20 minutes. After the annealing, the substrates are transferred to a nitrogen filled glovebox for the perovskite deposition. A stoichiometric perovskite solution (FAPbBr$_3$) is prepared by mixing 1 M PbBr$_2$ and 1 M FaBr in DMSO solvent. An ionic liquid precursor is added to the FAPbBr$_3$ solution before use. Then, the FAPbBr$_3$ perovskite films are deposited by spin coating at 4000 rpm for 20s using the solvent quenching method on warm substrates (60°C). Ethyl acetate (200 µl) is dropped after 10 seconds. The samples are then annealed at 80°C for 10 minutes. Neo-Pentylammonium chloride (NEO) and iso-Pentylammonium chloride (ISO) solutions are prepared at a 1 mg/ml concentration in 2-propanol and then mixed 1:1 v/v for 2D surface passivation. The mixed solution is deposited by spin coating at 4000 rpm for 20 s. The PTAA solution was prepared by dissolving 10 mg of the PTAA powder in 1 ml of toluene solvent and doped using TBP (10 µl/ml) and Li-TFSI (5 µl/ml, stock solution: 170 mg/ml in acetonitrile). The PTAA films are deposited by spin coating at 4000 rpm for 20 seconds. P2 ablation is performed by placing 6 lines beside different fluences of ~1 mJ cm$^{-2}$ difference, with an average fluence per pulse of 29.06 mJ cm$^{-2}$. Low temperature ITO deposition was performed by using an industrial in-line magnetron sputtering (KENOSISTEC S.R.L., KS 400 In-Line) at 1.1x10$^{-3}$ mbar and 90 W RF power. Inert Ar gas is purged in the chamber (40 sccm) during the ITO deposition to activate the Ar$^+$ plasma. The sample holder is moved below the ITO cathode at a 120 cm/min speed for 200 cycles to achieve 220 nm as thickness. The P3 ablation is performed by a single line with fluence per pulse of 21.23 mJ cm$^{-2}$. The module layout was scribed on a 60x60 mm$^2$ substrate, formed by 6 series-connected cells with 7.38 mm*45 mm cell width and height, 160 µm-wide P1-P2-P3 interconnection, 19.926 cm$^2$ active area and 97.83% geometrical fill factor. An antireflection coating of porous Al$_2$O$_3$ was deposited by spin coating a 5 wt% dispersion of Al$_2$O$_3$ nanoparticles in 2-propanol at 3000 rpm for 30s. After spin coating, the film was dried at 80°C for 2 minutes. The porosity of the film reduces the refractive index compared to bulk $Al_2O_3$ and provide a facile method to reduce the reflections at the ITO/air interface.

*Organic solar modules*

*Materials*

The inverted structure was chosen for the OPV modules in this work. The layer stack was glass/ITO/ZnO/PM6:Y6:$PC_{61}$BM/MoOx/Ag. 6 cm x 6 cm glass substrates coated with indium tin oxide (ITO) with a sheet resistance of 15 Ω/□ were purchased form VisionTek. Zinc Oxide (ZnO) nanoparticle dispersion (N-10;2.5 wt.-% in isopropanol) was purchased from Avantama. Poly[(2,6-(4,8-bis(5-(2-ethylhexyl-3-fluoro)thiophen-2-yl)-benzo[1,2-b:4,5-b´]dithiophene))-alt-(5,5-(1´,3´-di-2-thienyl-5´-7´-bis(2-ethylhexyl)bonzol[1´,2´-c:4´,5´-c´]dithiophene-4,8-dione)] (PBDB-T-2F or „PM6") and (2,2´-((2Z,2´Z)-(12,13-bis(2-ethylhexyl)-2,9-diundecyl-12,13-dihydro-[1,2,5]thiadiazolo[3,4-e]thieno[2",3":4´,5]thieno[2´,3´:4,5]pyrrolo[3,2-g]thienol[2´,3´:4,5]thieno[3,2b]indole-2,10-diyl)bis(methanylylidene))-bis(5,6-diflouro-3-oxo-2,3-dihydro-1H-indene-2,1-diylidene))dimalononitrile) („Y6") were purchased from Derthon OPV. [6,6]-phnyl-C61-butytic acid methyl ester ($PC_{61}$BM) was purchased from Nano-C. Silver (Ag) was purchased from Evochem.

*Device fabrication*

The laser patterning of the solar modules, "P1", "P2" and "P3", was conducted by means of a Spirit 1040-8 SHG laser from Spectra Physics ($P_{max}$= 6 W, λ =520 nm) with a femtosecond pulse width in combination with a galvanometer scanner with f-theta lens and a focal length of 506 mm. The module fabrication process was carried out as follows:

The glass/ITO substrates were "P1" laser patterned with the fs-laser, followed by a cleaning step with a microfiber tissue and toluene. N-10 was coated with a doctor blade and annealed for 30 minutes at 200 °C in ambient atmosphere. PM6:Y6:PC$_{61}$BM (1:1,1:0.1, 20 mg/ml in chloroform) was stirred at room temperature for 2 hours in a nitrogen-filled glovebox before being coated in air with a doctor blade using the following processing parameters: 120 µl of solution are injected into a 400 µm gap between substrate and blade (both heated to 30 °C). Coating with 5-10 mm/s formed a fast-drying wet film, that covered the substrate completely. The resulting dry films had a maximum optical density between 0.50 and 0.60. The devices were subsequently annealed at 110 °C for 10 minutes under inert atmosphere. Subsequently, 10-nm MoOx were thermally evaporated, followed by the "P2" laser patterning step. Finally, 50-nm Ag were thermally evaporated, followed by the "P3" laser patterning step.

The current–voltage characteristics of all the modules were recorded with a precision source measurement unit (B2901A by Keysight) under illumination by a class AAA solar simulator (LOT) providing an AM1.5G spectrum with an intensity of 100 mW/cm$^2$, which was calibrated by employing a silicon reference cell.

The tandem configuration of the perovskite and organic modules was realized by placing the organic solar module underneath the perovskite module, which was mounted on a glass cover by using stripes of sticky tape on the edges of the substrate far from the active area. The constituting cells of the semitransparent, top mini-module based on the wide band gap perovskite FAPbBr$_3$ were oriented perpendicularly with respect to those of the bottom organic module, as observed in the Figure 4b of the main text. In contrast to the parallel orientation, in this geometrical arrangement the light shining through the almost fully transparent interconnecting regions of the top perovskite module is equally distributed over the serially connected cells of the organic bottom sub-module.

During the I-V measurements, the performance of the individual as well as the tandem solar modules was recorded while keeping both devices inside a closed metallic box, in which a constant $N_2$ flow was maintained throughout the complete procedure. Sticky bus bars were attached to the module terminals to extract the photocurrent via wiring with electric clips, as seen in the Figure 4c.

The average and standard deviation of the photovoltaic parameters of the tandems were calculated over three consecutive I-V scans in reverse bias, whereas the measurements for the individual organic modules were performed in forward bias. The latter were exposed to light soaking for about 30 s before the beginning of the measurements.

To decrease the hysteresis and enhance the performance of the semitransparent wide bandgap perovskite modules, I-V scans in forward and reverse bias were recorded, followed by maximum power point tracking for 2 min and finally another round of I-V measurements in the same sequence were performed. When analyzed individually, a piece of black cardboard was placed at the bottom of the perovskite module (on the sputtered ITO electrode) to isolate its $I_{sc}$ from any light reflection coming from the metallic box. We observed a negligible impact on its $I_{sc}$ when the black cardboard was replaced by an organic module, as shown below.

**Supporting Information**

Supporting Information is available from the author.

**Author contributions**

J. G. C., A. D. and F. M. contributed equally. H.-J. E., C. J. B., and A. D. C. conceived the idea, supervised the project, revised and corrected the manuscript. J.G.C. made the filtering experiments, the preliminary module design for tandems, the tandem interconnection, the simulations in LTspice and wrote the manuscript. A. D. and R. B. fabricated the organic solar modules. A.D. performed the electrical simulations in COMSOL Multiphysics. F. M. fabricated the semitransparent perovskite modules. K. F. performed the optical simulations. L. A. C. optimized the lasering conditions for the perovskite modules. F. J. measured the optical properties of the films comprising the perovskite modules. M. W., F. B., F. Y., N. L., and A. N. C. M., revised and corrected the manuscript.

**Acknowledgments**

J.G.C. gratefully acknowledges the financial support of the Deutscher Akademischer Austauschdienst (DAAD) through a doctoral scholarship. The authors acknowledge the "Solar Factory of the Future" as part of the Energy Campus Nürnberg (EnCN), which is supported by the Bavarian State Government (FKZ 20.2-3410.5-4-5). We acknowledge funding from the European Union's Horizon 2020 research and innovation program under the grant agreement No. 101007084 (CITYSOLAR). H.-J. E. and C. J. B. acknowledge funding from the European Union's Horizon 2020 INFRAIA program under Grant Agreement No. 101008701 ('EMERGE'). Part of this work has been supported by the Helmholtz Association in the framework of the innovation platform "Solar TAP".

**Conflicts of Interest**

The authors declare no conflict of interest.

**Data Availability Statement**

The data that support the findings of this study are available from the corresponding authors upon reasonable request.

# Supporting Information

**Matching the photocurrent of perovskite/organic tandem solar modules by varying the cell width**

*José García Cerrillo\*, Andreas Distler, Fabio Matteocci, Karen Forberich, Michael Wagner, Robin Basu, Luigi Angelo Castriotta, Farshad Jafarzadeh, Francesca Brunetti, Fu Yang, Ning Li, Asiel Neftalí Corpus Mendoza, Aldo Di Carlo, Christoph J. Brabec\*, Hans-Joachim Egelhaaf*

J. García Cerrillo, Dr. A. Distler, R. Basu, Prof. N. Li, Dr. H.-J. Egelhaaf, Prof. C. J. Brabec
Institute of Materials for Electronics and Energy Technology (i-MEET)
Friedrich-Alexander-Universität Erlangen-Nürnberg (FAU)
Martensstraße 7, Erlangen, D-91058, Germany.
E-Mail address: jose.garcia@fau.de.

Dr. F. Matteocci, Dr. L. A. Castriotta, F. Jafarzadeh, Dr. Francesca Brunetti, Prof. A. Di Carlo
Center for Hybrid and Organic Solar Energy (CHOSE)
University of Rome Tor Vergata
via del Politecnico 1
00133 Rome, Italy.

Dr. K. Forberich, M. Wagner, Prof. N. Li, Dr. H.-J. Egelhaaf, Prof. C. J. Brabec
Helmholtz Institute Erlangen-Nürnberg for Renewable Energy (HI ERN)
Forschungszentrum Jülich GmbH
Immerwahrstraße 2, Erlangen, D-91058, Germany.
\* E-mail address: christoph.brabec@fau.de.

Dr. F. Yang
Laboratory of Advanced Optoelectronic Materials
College of Chemistry
Chemical Engineering and Materials Science
Soochow University
Suzhou 215123, China.

Prof. N. Li
State Key Laboratory of Luminescent Materials and Devices
Institute of Polymer Optoelectronic Materials and Devices
School of Materials Science and Engineering
South China University of Technology
Guangzhou 510640, China.

Dr. Asiel Neftalí Corpus Mendoza
Instituto de Energías Renovables
Universidad Nacional Autónoma de México
Priv. Xochicalco S/N, Temixco, Morelos, 62580 México.

Prof. A. Di Carlo
Institute of Structure of Matter (ISM)
National Research Council (CNR)
via del Fosso de Cavaliere 100
00133 Rome, Italy.

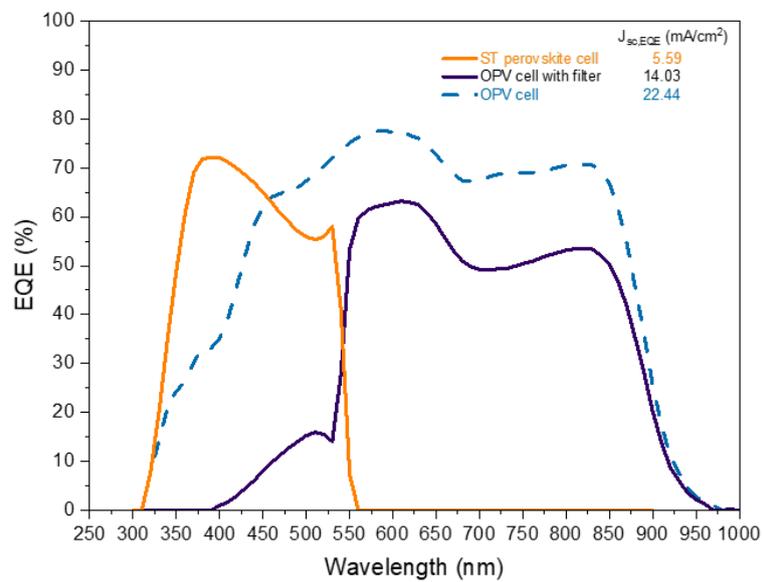

**Figure S1.** External Quantum Efficiency (EQE) spectra of semitransparent perovskite (FAPbBr$_3$) and organic (PM6:Y6:PCBM) solar cells.

# Efficiency simulations of tandem modules with equal and adjusted cell width

The efficiency estimates shown in Figure 3 are based on optical simulations performed with the transfer matrix model for the stacks as shown in the same figure. The optical constants used were obtained from our own measurements (ellipsometry or transmittance / reflectance) or taken from the literature and are shown in **Figure S2**.

(a)
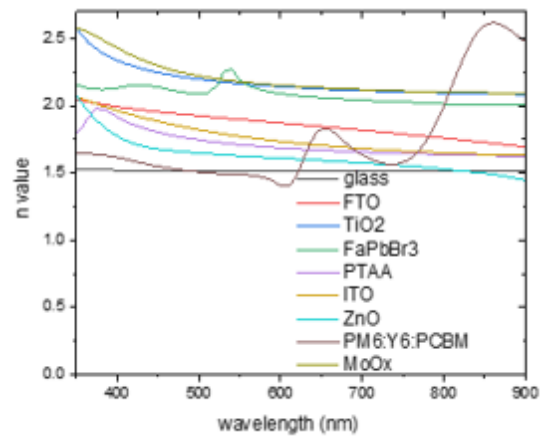

(b)
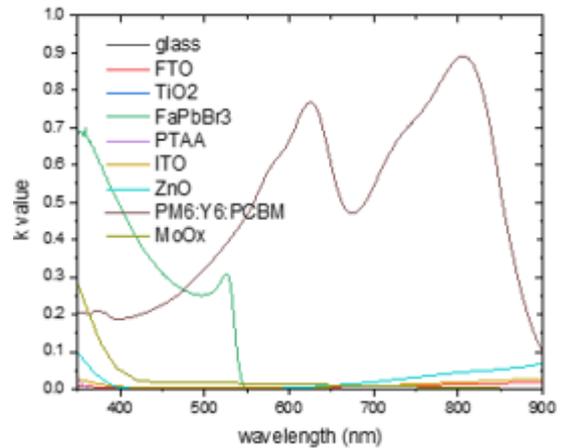

(c)
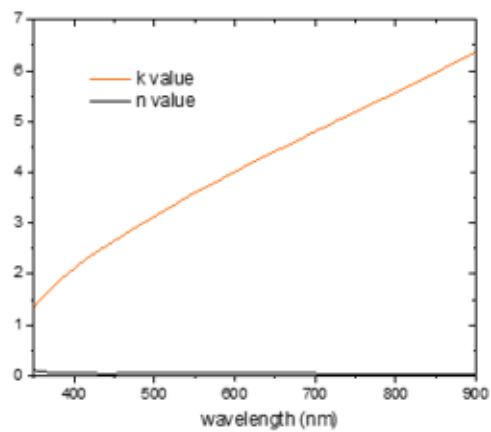

**Figure S2**. Optical constants used for the transfer matrix simulations. a) *n* values, b) *k* values, c) *n* & *k* values for silver (in a separate panel due to the different scale).

Specifically, the values for ITO and ZnO were determined by reflectance / transmittance measurements according to Čampa,[1] the values for PM6:Y6 were determined by ellipsometry, the *k* value for FAPbBr$_3$ was taken from transmittance measurements of the corresponding film, and all remaining values were taken from Rossi et al.[2]

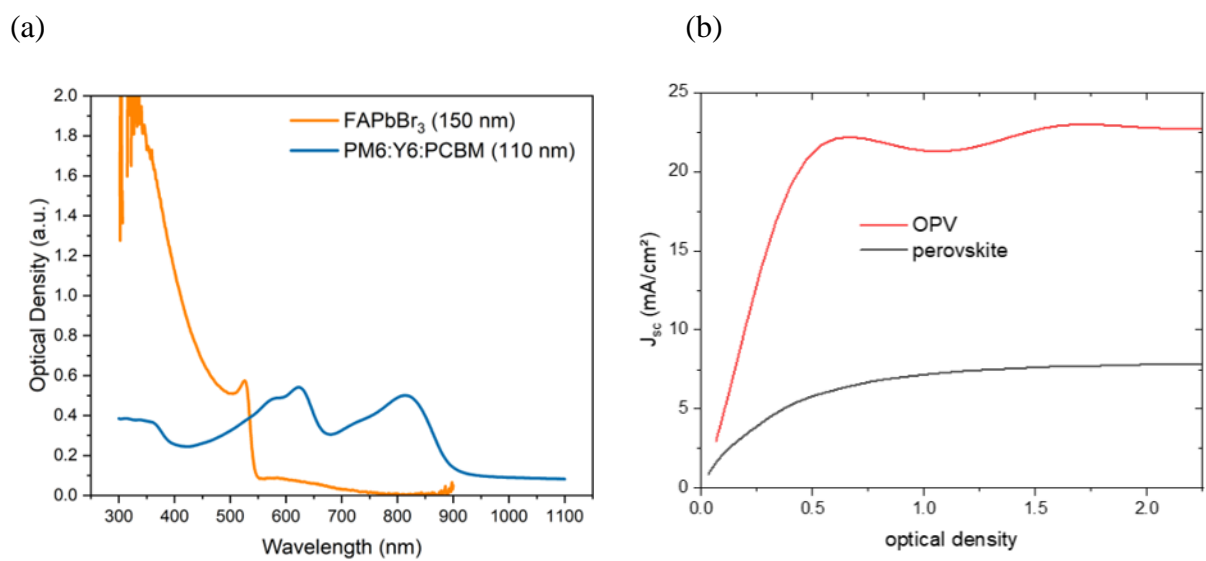

**Figure S3.** (a) Optical density (*OD*) spectra of the FAPbBr$_3$ and PM6:Y6:PCBM thin films, and (b) $J_{sc}$ values depending on layer thickness calculated for separate perovskite cell and non-shaded OPV cell.

The perovskite and organic modules were treated as separate thin film stacks, and the transmittance of the perovskite module is treated as the incident radiation for the OPV module. Also, the reflectance of the OPV module is incident on the perovskite module for a second time from the back (silver electrodes). $V_{oc}$ and *FF* values were taken from the measurements of single cells as $FF_{OPV}$ = 0.63, $FF_{PVKT}$ = 0.7, $V_{oc,OPV}$ = 0.8 V and $V_{oc,PVKT}$ = 1.5 V. A thickness dependence of these quantities was not considered in this approximation. The IQE values of OPV and perovskite were set as 80% and 100%, respectively, based on the comparison of the calculated with the measured $J_{sc}$ values of the single cells.

For the simulation of the tandem module with equal cell widths, we assumed the sum of the voltages and the smaller value of the current, the fill factor was taken as the average value of the single values. For the module with adjusted cell widths, the width of one cell of the OPV module was chosen so that the currents of both modules are equal, as illustrated in **Figure S4.** **Figure S4a** shows the ratio $J_{sc,f\text{-}OPV} / J_{sc,PVKT}$, and **Figure S4b** shows the resulting width of the cells in the OPV module for a perovskite cell width of 7.22 mm.

The higher $J_{sc}$ that is produced by an OPV cell compared to a perovskite cell thus leads to a larger number of cells and, therefore, to a higher $V_{oc}$ compared to the case of equal cell width.

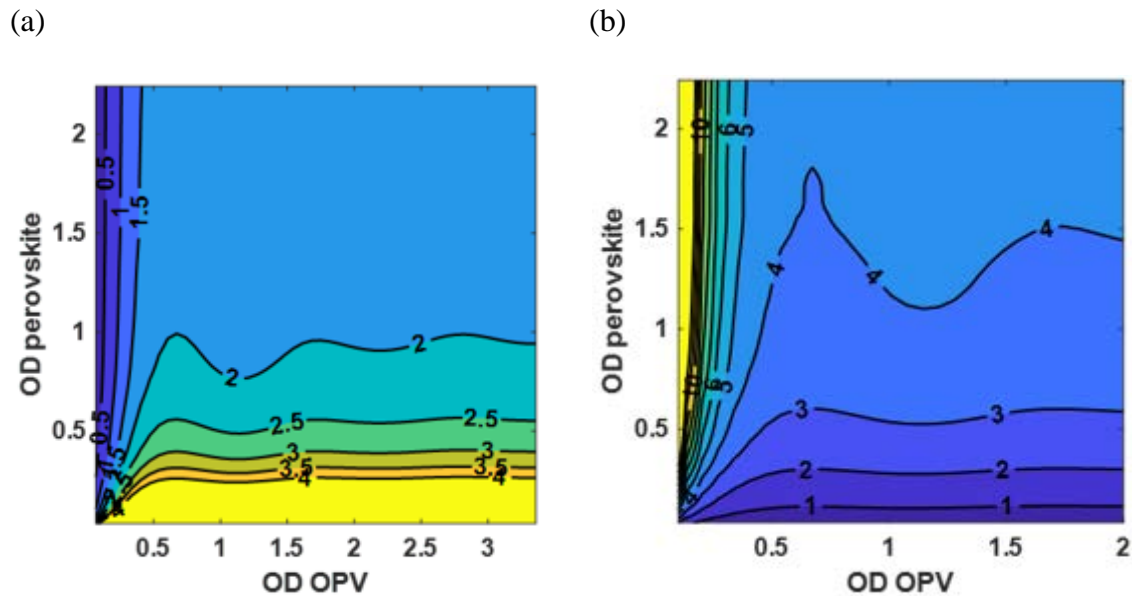

**Figure S4**. (a) Ratio of $J_{sc}$ in the OPV module to the $J_{sc}$ in the perovskite module and (b) resulting cell width (in mm) in the OPV module for a perovskite cell width of 7.22 mm.

Furthermore, for each cell width, additional efficiency losses in the organic modules due to series resistances (mainly in the ITO) and reduced active area (geometric fill factor) are considered according to **Figure S5**. This loss factor was obtained by electrical finite element method (FEM) simulations using COMSOL Multiphysics. The self-made model simulates the resulting *I-V* curve of a module as a function of cell number at fixed total module length, width and interconnect width. The resulting maximum power outputs are compared to that of an optimized cell, whose *I-V* curve is also used as input curve for the module simulation.

For the architecture with equal cell width, we assumed the minimum loss of 5% (assuming that the modules are fabricated with optimum cell width).

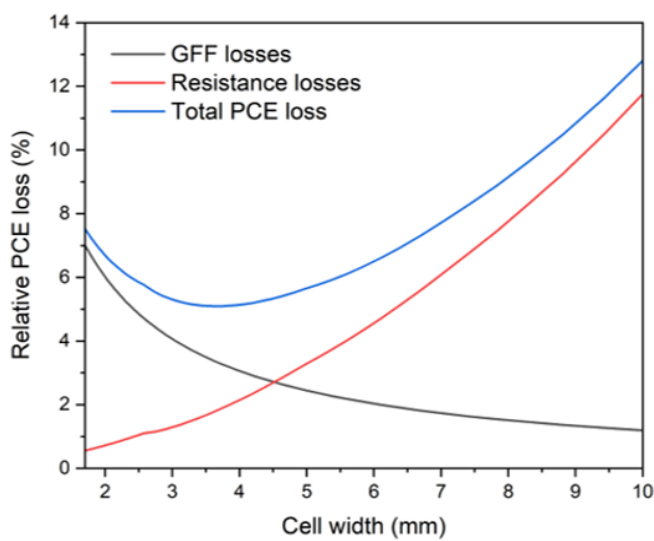

**Figure S5**. (a) Assumed loss factor depending on the width of one cell in the OPV module.

**Table S1.** Average and standard deviation of the parameters of the perovskite/organic tandem solar modules with varying sub-cell width or number of cells in the organic module.

| Module | | $J_{sc}$ (mA/cm$^2$) | $I_{sc}$ (mA) | $V_{oc}$ (V) | $I_{mpp}$ (mA) | $V_{mpp}$ (V) | $P_{mpp}$ (mW) | FF (%) | $PCE_{mod}$ (%) |
|---|---|---|---|---|---|---|---|---|---|
| **OPV with 12 cells (3.63 mm)** | | | | | | | | | |
| ST-Perovskite | | 6.15 ± 0.01 | 20.4 ± 0.02 | 9.11 ± 0.04 | 17.14 ± 0.18 | 7.13 ± 0.05 | 122.16 ± 1.70 | 65.76 ± 0.62 | **6.00 ± 0.08** |
| Organic (filtered) | | 13.59 ± 0.02 | 22.2 ± 0.03 | 8.87 ± 0.00 | 19.15 ± 0.09 | 6.78 ± 0.00 | 129.78 ± 0.63 | 65.86 ± 0.33 | 6.41 ± 0.03 |
| Organic | | 22.43 ± 0.03 | 36.65 ± 0.05 | 9.55 ± 0.00 | 31.55 ± 0.24 | 7.03 ± 0.05 | 221.85 ± 0.54 | 63.41 ± 0.03 | **10.96 ± 0.03** |
| Tandem | Series | 1.06 ± 0.00 | 21.5 ± 0.00 | 18.09 ± 0.01 | 18.05 ± 0.21 | 14.18 ± 0.08 | 255.9 ± 1.73 | 65.81 ± 0.33 | **12.64 ± 0.09** |
| | Parallel | 2.12 ± 0.05 | 42.96 ± 0.08 | 9.00 ± 0.01 | 36.73 ± 0.34 | 6.94 ± 0.05 | 254.72 ± 0.68 | 65.9 ± 0.08 | **12.58 ± 0.03** |
| **OPV with 13 cells (3.19 mm)** | | | | | | | | | |
| ST-Perovskite | | 6.46 ± 0.01 | 21.45 ± 0.03 | 9.18 ± 0.02 | 18.14 ± 0.17 | 7.1 ± 0.09 | 128.68 ± 0.60 | 65.39 ± 0.13 | **6.32 ± 0.03** |
| Organic (filtered) | | 14.6 ± 0.02 | 20.94 ± 0.03 | 10.63 ± 0.00 | 18.06 ± 0.19 | 8.36 ± 0.17 | 150.91 ± 1.48 | 67.81 ± 0.79 | 7.45 ± 0.07 |
| Organic | | 22.95 ± 0.02 | 32.91 ± 0.03 | 10.79 ± 0.01 | 28.57 ± 0.30 | 8.81 ± 0.10 | 251.67 ± 0.59 | 70.88 ± 0.23 | **12.43 ± 0.03** |
| Tandem | Series | 1.04 ± 0.00 | 21.11 ± 0.00 | 19.89 ± 0.03 | 18.05 ± 0.02 | 15.9 ± 0.00 | 286.95 ± 0.24 | 68.34 ± 0.15 | **14.17 ± 0.01** |
| | Parallel | 2.13 ± 0.07 | 43.08 ± 0.09 | 10.05 ± 0.01 | 36.99 ± 0.44 | 7.63 ± 0.08 | 282.23 ± 0.36 | 65.17 ± 0.10 | **13.94 ± 0.02** |

(a)

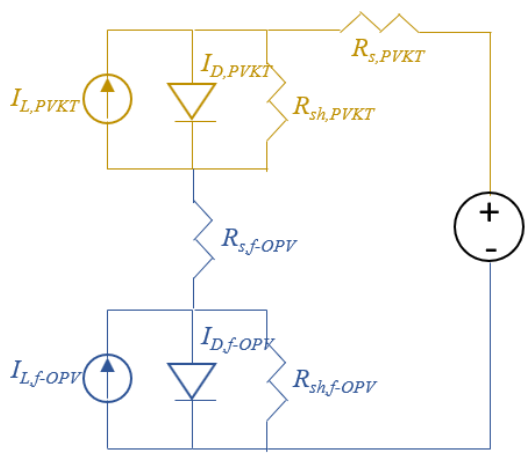

For the individual sub-modules:

$$I = I_0\left\{\exp\left[\frac{q(V-IR_s)}{n_a k_B T}\right] - 1\right\} + \frac{V-IR_s}{R_{sh}} - I_L$$

(b)

|  | $R_{sh}= 10^4$ kΩ | $R_{sh}= 10^3$ kΩ |
|---|---|---|
| $I_{sc,PVKT}$ | 21.23 mA | 21.23 mA |
| $I_{sc,tandem}$ | 21.18 mA | 21.19 mA |
| $I_{sc,f-OPV}$ | 21.18 mA | 21.18 mA |
|  | $R_{sh}= 10^4$ kΩ | $R_{sh}= 10^3$ kΩ |

(c)

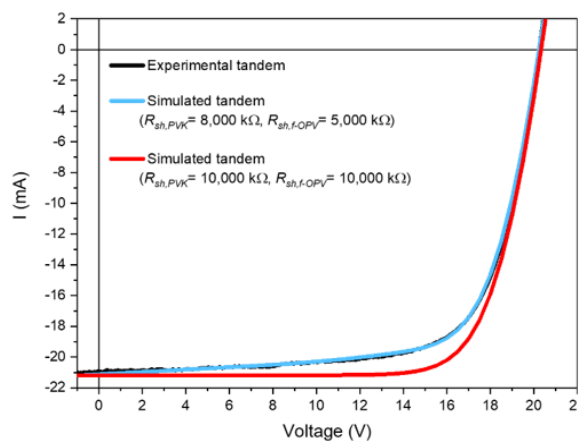

**Figure S6.** (a) Equivalent circuit of a perovskite/organic tandem module, (b) Relative position of the tandem $I_{sc}$ between the individual $I_{sc}$ at two different shunt resistances, $10^3$ and $10^4$ kΩ, and (c) the corresponding experimental and simulated *I-V* curves.

**Table S2.** Parameters used for the simulations in LTspice. For both the perovskite and filtered organic sub-modules, (a) $R_{sh}= 10^3$ k$\Omega$, and (b) $R_{sh}= 10^4$ k$\Omega$.

(a)

| Parameter | Perovskite | Filtered OPV | Tandem |
|---|---|---|---|
| $I_L$ (A) | 21.23 | 21.18 | |
| $I_0$ (A) | $1\times10^{-10}$ | $1\times10^{-10}$ | |
| $n_a$ | 19 | 22 | |
| $R_s$ ($\Omega$) | 17 | 37 | |
| $R_{sh}$ (k$\Omega$) | $10^3$ | $10^3$ | |
| $I_{sc}$ (mA) | 21.23 | 21.18 | **21.19** |
| $V_{oc}$ (V) | 9.41 | 10.90 | 20.33 |
| FF (%) | 76.46 | 73.77 | 75.26 |
| PCE (%) | 7.54 | 8.41 | 16.00 |

(b)

| Parameter | Perovskite | Filtered OPV | Tandem |
|---|---|---|---|
| $I_L$ (A) | 21.23 | 21.18 | |
| $I_0$ (A) | $1\times10^{-10}$ | $1\times10^{-10}$ | |
| $n_a$ | 19 | 22 | |
| $R_s$ ($\Omega$) | 17 | 37 | |
| $R_{sh}$ (k$\Omega$) | $10^4$ | $10^4$ | |
| $I_{sc}$ (mA) | 21.23 | 21.18 | **21.18** |
| $V_{oc}$ (V) | 9.41 | 10.90 | 20.33 |
| FF (%) | 76.48 | 73.80 | 75.30 |
| PCE (%) | 7.83 | 8.42 | 16.01 |

**Table S3**. Brief summary of recently reported 2-terminal perovskite tandem solar cells with a small $J_{sc}$ mismatch.

| Reference | $J_{sc}$ (mA/cm²) | | | $\Delta J_{sc}$* (%) |
|---|---|---|---|---|
| | **Top** | **Bottom** | **Tandem** | |
| 3 | 9.13 | 9.29 | 9.7 | 4.41 |
| 4 | 13.79 | 13.75 | 13.92 | 1.24 |
| 5 | 18.75 | 18.64 | 18.78 | 0.75 |
| 6 | 17.9 | 14.7 | 15.3 | 4.08 |
| 7 | 14.6 | 13.3 | 13.8 | 3.76 |
| 8 | 19.84 | 20.02 | 19.35 | -3.35 |
| 9 | 16.7 | 16.8 | 16.5 | -1.79 |
| 10 | 14.98 | 12.36 | 13.6 | 10.03 |
| 11 | 15.1 | 12.6 | 13.1 | 3.97 |
| 12 | 15.7 | 14.4 | 15.1 | 4.86 |

* $\Delta J_{sc}$ is defined as $(J_{sc,tandem} - J_{sc,bottom})*100/J_{sc,bottom}$. The reported top and bottom $J_{sc}$ values were obtained by measuring the sub-cell external quantum efficiency (EQE), whereas the tandem $J_{sc}$ was obtained via measuring the device I-V characteristics.

**Table S4**. Experimental short circuit current of the FAPbBr$_3$ module before and after placing a PM6:Y6:PCBM module underneath.

| Back-reflection from the OPV module | $J_{sc}$ (mA/cm²) | $I_{sc}$ (mA) |
|---|---|---|
| No | 6.45 | 21.41 |
| Yes | 6.66 | 22.10 |

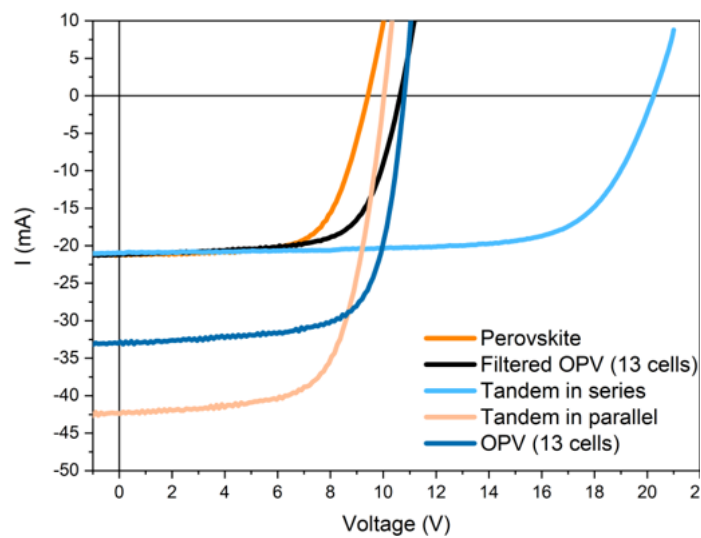

**Figure S7.** Current-voltage characteristics of a perovskite/organic tandem using a perovskite module with enhanced power conversion efficiency.

**Table S5.** Photovoltaic parameters of a perovskite/organic tandem in series using a perovskite module with enhanced power conversion efficiency.

| Module | | $J_{sc}$ (mA/cm²) | $I_{sc}$ (mA) | $V_{oc}$ (V) | $I_{mpp}$ (mA) | $V_{mpp}$ (V) | $P_{mpp}$ (mW) | FF (%) | $PCE_{mod}$ (%) |
|---|---|---|---|---|---|---|---|---|---|
| ST-Perovskite | | 6.40 | 21.23 | 9.41 | 18.43 | 7.35 | 135.53 | 67.83 | **6.65** |
| Organic (filtered) | | 14.77 | 21.18 | 10.64 | 18.14 | 8.46 | 153.51 | 68.15 | 7.58 |
| Organic | | 22.98 | 32.97 | 10.79 | 28.76 | 8.78 | 252.41 | 70.95 | **12.46** |
| Tandem | Experimental | 1.03 | 20.94 | 20.23 | 18.30 | 16.40 | 300.07 | 70.82 | **14.82** |
| | Expected | 1.05 | 21.18 | 20.05 | | | 288.70 | 67.99 | 14.26 |

**Table S6.** Average and standard deviation of the photovoltaic parameters of a perovskite/organic tandem in series with enhanced power conversion efficiency.

| Module | $J_{sc}$ (mA/cm²) | $I_{sc}$ (mA) | $V_{oc}$ (V) | $I_{mpp}$ (mA) | $V_{mpp}$ (V) | $P_{mpp}$ (mW) | FF (%) | $PCE_{mod}$ (%) |
|---|---|---|---|---|---|---|---|---|
| ST-Perovskite | 6.36 ± 0.03 | 21.11 ± 0.08 | 9.42 ± 0.01 | 18.26 ± 0.12 | 7.42 ± 0.05 | 135.41 ± 0.10 | 68.1 ± 0.19 | **6.65 ± 0.00** |
| Organic (filtered) | 14.75 ± 0.01 | 21.16 ± 0.02 | 10.64 ± 0.00 | 18.04 ± 0.07 | 8.43 ± 0.05 | 152.1 ± 1.26 | 67.56 ± 0.56 | 7.51 ± 0.06 |
| Organic | 20.14 ± 0.04 | 32.91 ± 0.06 | 10.80 ± 0.01 | 28.45 ± 0.33 | 8.85 ± 0.10 | 251.57 ± 0.65 | 70.77 ± 0.13 | **12.42 ± 0.03** |
| Tandem | 1.05 ± 0.00 | 21.30 ± 0.00 | 20.19 ± 0.02 | 18.39 ± 0.13 | 16.19 ± 0.08 | 297.75 ± 0.82 | 69.24 ± 0.26 | **14.70 ± 0.04** |

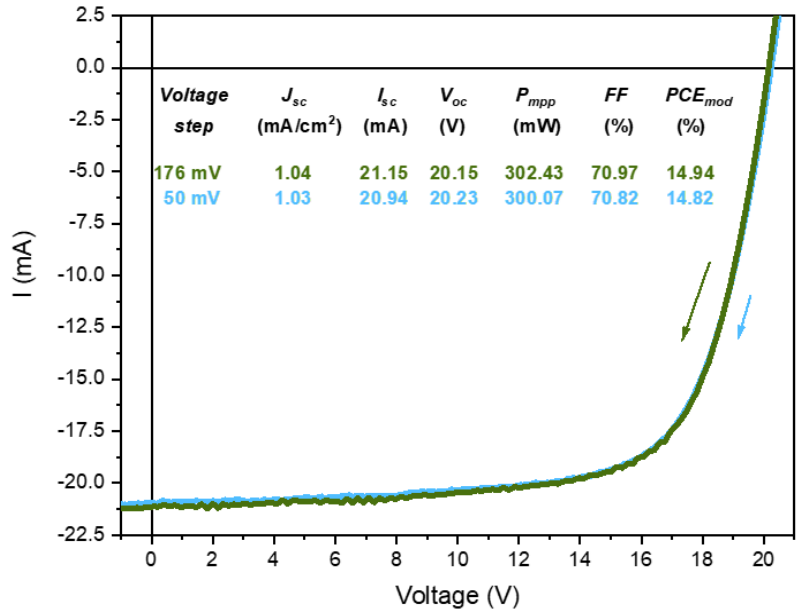

**Figure S8.** Current-voltage characteristics of the champion perovskite/organic tandem solar module at different voltage step sizes.

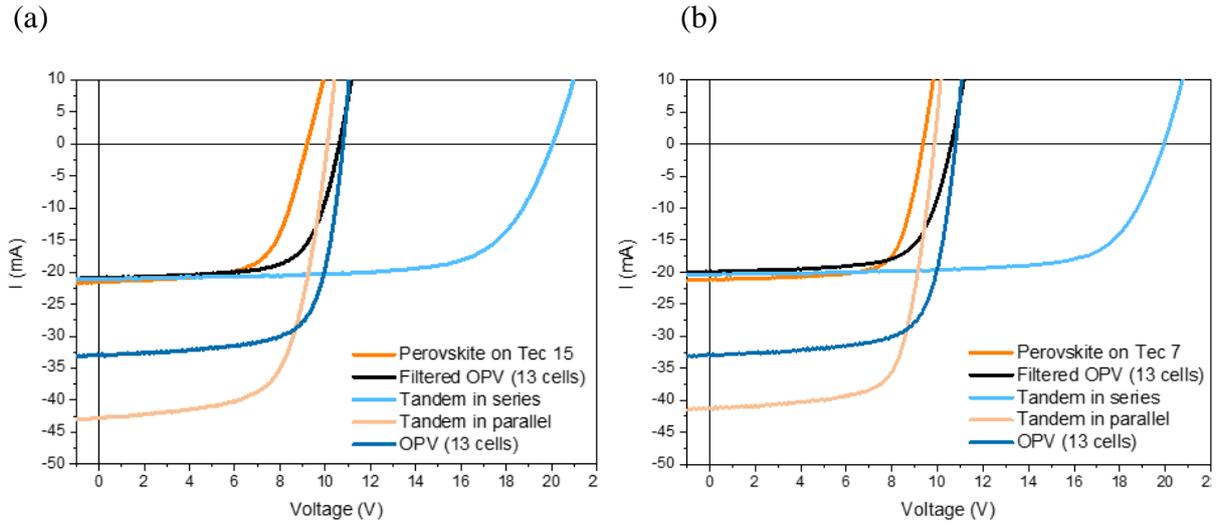

**Figure S9.** Current-voltage characteristics of tandem perovskite/organic solar modules using FTO (a) Tec 15 and (b) Tec 7 substrates to fabricate the perovskite modules.

**Table S7.** Photovoltaic parameters of perovskite/organic tandems in series using FTO Tec 15 and Tec 7 substrates for the perovskite modules.

| Module | | $J_{sc}$ (mA/cm²) | $I_{sc}$ (mA) | $V_{oc}$ (V) | $I_{mpp}$ (mA) | $V_{mpp}$ (V) | $P_{mpp}$ (mW) | FF (%) | $PCE_{mod}$ (%) |
|---|---|---|---|---|---|---|---|---|---|
| **Tec 15** | | | | | | | | | |
| ST-Perovskite | | 6.46 | 21.46 | 9.20 | 18.08 | 7.16 | 129.45 | 65.55 | **6.36** |
| Organic (filtered) | | 14.57 | 20.90 | 10.62 | 17.80 | 8.57 | 152.54 | 68.68 | 7.53 |
| Organic | | 22.92 | 32.87 | 10.79 | 28.43 | 8.88 | 252.50 | 71.20 | **12.47** |
| Tandem | Experimental | 1.04 | 21.07 | 20.05 | 17.93 | 16.20 | 290.48 | 68.76 | **14.34** |
| | Expected | 1.03 | 20.90 | 19.83 | | | 278.16 | 67.12 | 13.74 |
| **Tec 7** | | | | | | | | | |
| ST-Perovskite | | 6.36 | 21.11 | 9.36 | 18.72 | 7.64 | 143.01 | 72.37 | **7.02** |
| Organic (filtered) | | 13.90 | 19.94 | 10.61 | 17.17 | 8.46 | 145.30 | 68.70 | 7.18 |
| Organic | | 22.92 | 32.87 | 10.79 | 28.43 | 8.88 | 252.50 | 71.20 | **12.47** |
| Tandem | Experimental | 1.01 | 20.38 | 19.94 | 17.48 | 16.55 | 289.25 | 71.16 | **14.28** |
| | Expected | 0.98 | 19.94 | 19.97 | | | 280.83 | 70.53 | 13.87 |

**Table S8.** Average and standard deviation of the photovoltaic parameters of perovskite/organic tandems (in series) using FTO Tec 15 and Tec 7 substrates for the perovskite modules.

| Module | $J_{sc}$ (mA/cm$^2$) | $I_{sc}$ (mA) | $V_{oc}$ (V) | $I_{mpp}$ (mA) | $V_{mpp}$ (V) | $P_{mpp}$ (mW) | FF (%) | $PCE_{mod}$ (%) |
|---|---|---|---|---|---|---|---|---|
| **Tec 15** | | | | | | | | |
| ST-Perovskite | 6.46 ± 0.01 | 21.45 ± 0.03 | 9.18 ± 0.02 | 18.14 ± 0.17 | 7.1 ± 0.09 | 128.68 ± 0.60 | 65.39 ± 0.13 | **6.32 ± 0.03** |
| Organic (filtered) | 14.60 ± 0.02 | 20.94 ± 0.03 | 10.63 ± 0.00 | 18.06 ± 0.19 | 8.36 ± 0.17 | 150.91 ± 1.48 | 67.81 ± 0.79 | 7.45 ± 0.07 |
| Organic | 22.95 ± 0.02 | 32.91 ± 0.03 | 10.79 ± 0.01 | 28.57 ± 0.30 | 8.81 ± 0.10 | 251.67 ± 0.59 | 70.88 ± 0.23 | **12.43 ± 0.03** |
| Tandem | 1.04 ± 0.00 | 21.11 ± 0.00 | 19.89 ± 0.03 | 18.05 ± 0.02 | 15.9 ± 0.00 | 286.95 ± 0.24 | 68.34 ± 0.15 | **14.17 ± 0.01** |
| **Tec 7** | | | | | | | | |
| ST-Perovskite | 6.34 ± 0.01 | 21.06 ± 0.04 | 9.34 ± 0.01 | 18.69 ± 0.15 | 7.61 ± 0.05 | 142.15 ± 0.73 | 72.23 ± 0.21 | **6.98 ± 0.04** |
| Organic (filtered) | 13.88 ± 0.02 | 19.90 ± 0.03 | 10.61 ± 0.00 | 17.1 ± 0.07 | 8.4 ± 0.10 | 143.54 ± 1.59 | 67.98 ± 0.64 | 7.09 ± 0.08 |
| Organic | 22.95 ± 0.02 | 32.91 ± 0.03 | 10.79 ± 0.01 | 28.57 ± 0.30 | 8.81 ± 0.10 | 251.67 ± 0.59 | 70.88 ± 0.23 | **12.43 ± 0.03** |
| Tandem | 1.01 ± 0.02 | 20.40 ± 0.03 | 19.78 ± 0.01 | 17.72 ± 0.15 | 16.25 ± 0.14 | 287.85 ± 0.24 | 71.35 ± 0.12 | **14.22 ± 0.01** |

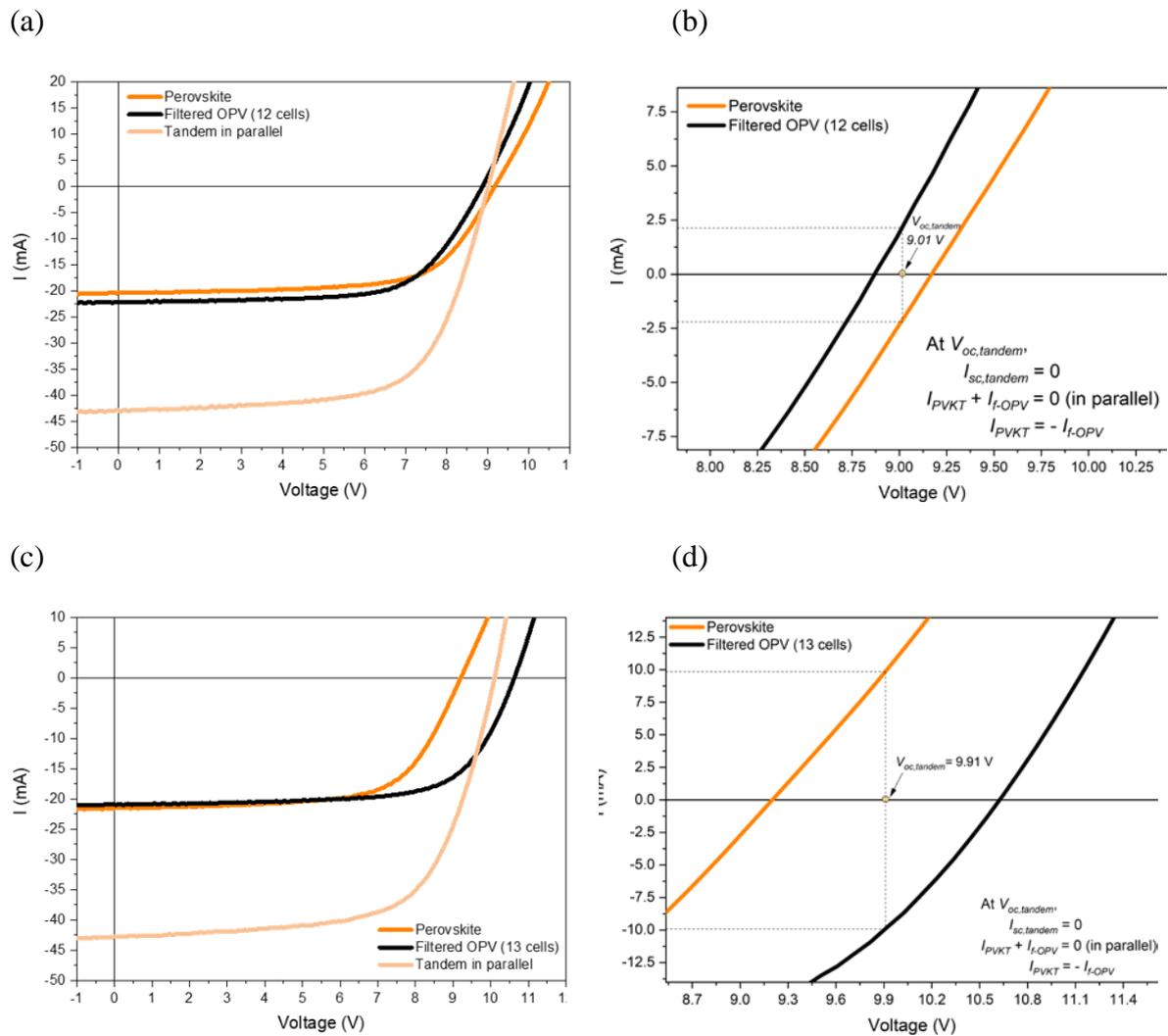

**Figure S10**. Current-voltage curves of individual perovskite and (a-b) 12 cell-, and (c-d) 13-cell-organic modules. Figures (b) and (d) show the geometric constructions employed to estimate the tandem $V_{oc}$.

For a parallel interconnection, the total current delivered by the tandem ($I_{tandem}$) equals the sum of the individual currents delivered by the perovskite ($I_{PVKT}$) and shaded organic modules ($I_{f\text{-}OPV}$) at a given voltage. However, when the applied voltage equals the tandem $V_{oc}$, no net current can be extracted from the device, which means that $I_{tandem} = I_{PVKT} + I_{f\text{-}OPV} = 0$. From this, it follows that $I_{PVKT} = -I_{f\text{-}OPV}$. In the region between the perovskite and shaded organic $V_{oc}$, there will be a point on the $V$ axis ($I= 0$) that lies on the straight line uniting the same $I$ value on the individual curves but of opposite sign; this value represents the expected tandem $V_{oc}$, which is equal to 9.01 V and 9.91 V for the tandems built with the organic modules having 12 and 13 cells, respectively. A similar procedure can be used to estimate the tandem $I_{sc}$, the difference is to make the geometric construction in the vicinity of the individual $I_{sc}$. In this case, $V_{tandem} = V_{PVKT} + V_{f\text{-}OPV} = 0$ on the $I$ axis and, therefore, $V_{PVKT} = -V_{f\text{-}OPV}$. The tandem $I_{sc}$ will lie on the line passing through the same voltage values but of opposite sign.